\documentclass[final,1p,times]{elsarticle} % do not change
\usepackage{graphicx} % do not change
\usepackage{amssymb} % do not change
\usepackage{amsthm} % do not change
\usepackage{lineno} % do not change

\usepackage{amsmath}
\usepackage{units}

\renewcommand{\d}{\text{d}}

\journal{Nuclear Physics A} % do not change
\begin{document} % do not change

\begin{frontmatter} % do not change

%% QM09Author: please enter your  
%% Title, author and address info here; please do not use footnotes

% Your Title - please insert
\title{LPM-Effect in Monte Carlo Models of Radiative Energy Loss}

% Principle author, and co-authors - please insert
\author{Korinna C. Zapp$^{a,b}$, Johanna Stachel$^{a}$, Urs Achim
Wiedemann$^{c}$}

% Address - please insert
\address[a]{Physikalisches Institut, Universit\"at Heidelberg, Philosophenweg
12, D-69120 Heidelberg, Germany}
\address[b]{ExtreMe Matter Institute EMMI, GSI Helmholtzzentrum
f\"ur Schwerionenforschung GmbH, Planckstra\ss e 1, 64291 Darmstadt, Germany}
\address[c]{Physics Department, Theory Unit, CERN, CH-1211 Gen\`eve 23,
Switzerland}

\begin{abstract} % do not change

Extending the use of Monte Carlo (MC) event generators to jets in
nuclear collisions requires a probabilistic implementation of the
non-abelian LPM effect. We demonstrate that a local, probabilistic
MC implementation based on the concept of formation times can account
fully for the LPM-effect. The main features of the analytically known
eikonal and collinear approximation can be reproduced, but we show how
going beyond this approximation can lead to qualitatively different
results.

\end{abstract} % do not change

\end{frontmatter} % do not change

%% QM09: we keep linenumbers at least for initial version
%\linenumbers % do not change

\section{Introduction}

With \textsc{Jewel} (Jet Evolution With Energy Loss) we work towards a
dynamically consistent Monte Carlo event generator (MC) for jet quenching that
is consistent with all analytically known limiting cases. These limiting
cases are the parton shower evolution in the absence of medium effects (as in
$e^+e^-$ or $pp$ collisions), energy loss due to elastic scattering in a medium
in the absence of vacuum radiation, medium induced gluon emission (radiative
energy loss) in the absence of vacuum radiation and to some extent induced
gluon emission in the presence of vacuum radiation. The latter is the only part
that is not fully implemented in \textsc{Jewel} yet.

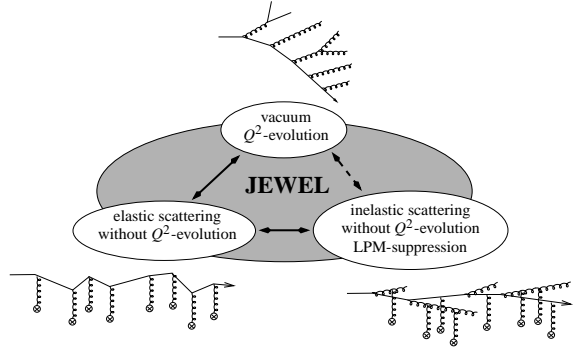
\begin{figure}[h!]
\centering
\input{qm09-jewel-diagram3_proc.pstex_t}\par
\end{figure}

For a detailed discussion of the vacuum parton shower, elastic energy loss and
the combination of both we refer to~\cite{Zapp:2008gi}. Here, we focus on the
Monte Carlo implementation of medium induced gluon bremsstrahlung,
where it is known from analytical calculations that a
quantum mechanical interference , namely the non-abelian
Landau-Pomerantschuk-Migdal (LPM) effect,
plays an important role. 

\section{Reproducing the BDMPS-ASW Results}
\label{sec:limcase}

Inspection of the field theoretical
calculations~\cite{Wiedemann:2000za,Baier:1996sk} reveals that radiated gluons
have a formation time
\begin{equation}
t_\text{f}=\frac{2 \omega}{k_\perp^2} \,,
\end{equation}
that can be interpreted as the time it takes the gluon to decohere from the
radiating projectile, i.e.\ to build up a relative phase that is of order
unity. During this time the gluon may acquire additional transverse momentum due
to multiple elastic rescattering, but further gluon emission off the
projectile
is suppressed. This suggests that the interference can be taken into account by
a MC algorithm that correctly treats the formation time
effects~\cite{Zapp:2008af}. To be more specific, the position of the inelastic
process is determined probabilistically and a gluon is produced with
energy
$\omega$ and transverse momentum $k_\perp$. Then the position of the next
interaction is chosen. If the distance to this scattering centre is
larger than
the gluon formation time $t_\text{f}$, then the gluon is regarded as fully
formed independent particle and the scattering centre can be the source of
further gluon radiation. If, on the other hand, the scattering centre is found
within the formation time the momentum transfer $q_\perp$ is added
coherently to
the gluon production process. This changes the formation time. If now the next
scattering occurs outside the updated formation time the gluon is regarded as
formed, otherwise the procedure is repeated. 

\begin{figure}[ht]
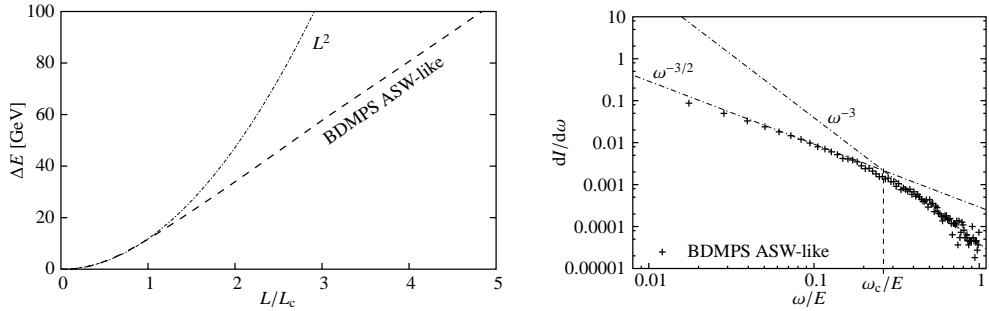

\centering
\input{qm09-deltae-bdmps_proc.pstex_t}\hspace{.5cm}
\input{qm09-didomega-2b_proc.pstex_t}\par
\caption[]{LHS: Total energy loss due to gluon radiation as a function of the
path length $L$ in the medium. RHS: Gluon energy spectrum at $L=L_\text{c}/2$.
'BDMPS ASW-like' denotes a setup of the MC simulation that parallels as much as
possible the assumptions of the analytical BDMPS ASW calculation. The
parameter choice corresponds to $\hat q \approx \unit[1]{GeV^2/fm}$, the
projectile energy is $E=\unit[100]{GeV}$.}
\label{fig:deltae_bdmps}
\end{figure}

The analytical calculations~\cite{Wiedemann:2000za,Baier:1996sk} are
carried out in the kinematical regime
$E \gg \omega \gg k_\perp, q_\perp > \Lambda_\text{QCD}$.
In order to reproduce the results of these calculations these
approximations
have to be respected by the MC. As a consequence, the energy degradation of the
projectile is neglected, the medium transfers only transverse momentum to the 
gluon, the momentum transfer is soft, the scattering centre takes no
recoil and the gluon $k_\perp$ in the initial production process has to be
small. We will discuss the consequences of these approximations in
more detail in section~\ref{sec:beyondlimcase}.

Characteristics for the BDMPS-ASW
calculation~\cite{Wiedemann:2000za,Baier:1996sk} are the modification of the
gluon spectrum and the path length dependence of the energy loss. The gluon
spectrum, which follows $\d I/\d\omega \propto \omega^{-1}$ for incoherent
production, is modified to $\d I/\d\omega \propto \omega^{-3/2}$ for $\omega <
\omega_\text{c}$ and $\omega^{-3}$ for $\omega >\omega_\text{c}$. The
characteristic gluon energy $\omega_\text{c} \simeq \hat q L^2/4$ is the
highest gluon energy that can be radiated coherently in a medium of length $L$.
The energy loss $\Delta E$ increases quadratically with $L$ for $L<L_\text{c}$,
for $L>L_\text{c}$ it becomes linear, $L_\text{c}\simeq \sqrt{4 E/\hat q}$ is
the formation time corresponding to the highest kinematically allowed
gluon energy. 

Figure~\ref{fig:deltae_bdmps} shows the results of our MC simulation
for the
energy loss and the gluon spectrum in the coherent regime at $L=L_\text{c}/2$.
The results show the features expected from the analytical BDMPS-ASW
calculation.

\section{Going Beyond the BDMPS-ASW Approximation}
\label{sec:beyondlimcase}

The MC simulation is a tool that allows one to go beyond the
approximations
discussed in the previous section. We successively relax the assumptions to
arrive at a more realistic description of radiative energy loss. The steps are

\begin{enumerate}
 \item \textit{relax soft gluon approximation:} the full power law
tail of the
elastic cross section is taken into account
\[\frac{\d\sigma}{\d q_\perp^2} \propto \frac{1}{(q_\perp^2+\mu^2)^2}\,
\theta(q_\perp^2-4\mu^2) \qquad \to \qquad
\frac{1}{(q_\perp^2+\mu^2)^2}\]
 \item \textit{energy conservation:} the degradation of the projectile
energy
due to gluon radiation is accounted for
 \item \textit{realistic kinematics of inelastic process:} the gluon is
produced with a transverse momentum $k_\perp$ distributed as $1/k_\perp^2$ (as
suggested, for instance, by full inelastic matrix elements); without the full
kinematics introduced in the next step this may be seen as a rather extreme
choice
 \item \textit{recoil:} the scattering centre in elastic and
inelastic scattering events becomes dynamic and takes recoil; to arrive at a
\textsc{Jewel}-like scenario elastic scattering of the projectile also has
to
be included
\end{enumerate}

\begin{figure}[ht]
\centering
\input{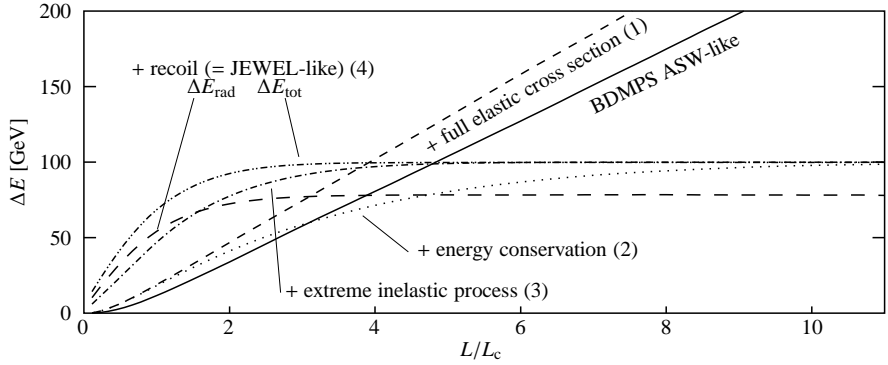}\par
\caption[]{Energy loss as function of in-medium path length for the stepwise
relaxation of assumptions. In the \textsc{Jewel} scenario a distinction
between radiative energy loss and total energy loss (including the energy taken
by recoiling scattering centres) has to be made.}
\label{fig:deltae_jewel}
\end{figure}

\begin{figure}[ht]
\centering
\input{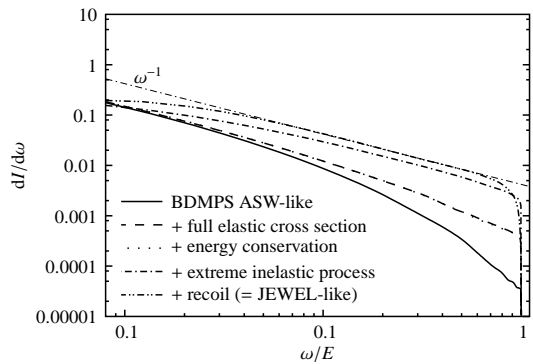}\par
\caption[]{Gluon energy spectrum at $L=L_\text{c}/2$ for the stepwise
relaxation of assumptions. Note that energy conservation plays only a minor
role in this regime, so that the third lines lies on top of the second.}
\label{fig:didomega}
\end{figure}

The parameter $\mu$ and
the mean free path $\lambda$ are kept fixed, only in the \textsc{Jewel}
scenario the elastic cross section is characterised by the infra-red
regulator and the momentum distribution of the scattering centres. The
energy loss and gluon
spectrum for the different stages are shown in
figures~\ref{fig:deltae_jewel} and \ref{fig:didomega}. Energy conservation
turns out to be an important effect $L>L_\text{c}$, but plays a negligible role
in the coherent regime for $L<L_\text{c}$. Generally, the radiation looks more
 and more incoherent. The modifications of the elastic and inelastic cross
sections lead to a loss of coherence. This is due to fluctuations that allow for
much smaller formation times, i.e.\ large momentum transfers in elastic
rescattering or large gluon $k_\perp$ in the inelastic process. The
approximation
$k_\perp \ll \omega$ in the BDMPS ASW calculation leads to long gluon formation
times after the inelastic process and the soft scattering approximation inhibits
a fast decoherence. In that sense the analytical calculation ensures
maximal
LPM-suppression, which can perhaps partly explain the large values of $\hat q$
that seem to be necessary to describe RHIC jet quenching data.

\section{Outlook}

We discussed how medium induced gluon radiation showing LPM interference can be
implemented in a MC event generator without modifying the gluon
emission process. This is an important step towards a dynamically
consistent
implementation of radiative energy loss. However, in order to arrive at a
complete description of jet evolution in dense QCD matter, the
algorithm has
to be generalised to induced gluon emission in the presence of vacuum
radiation. In this case there is interference between vacuum and medium
induced radiation in addition to the interference between subsequent
induced
emissions. An appropriate MC algorithm generalising the prescription discussed
here can be constructed and will be implemented in \textsc{Jewel}. This will
complete the description of interactions with the medium in \textsc{Jewel}.

%% end of main text

\section*{Acknowledgments} 
This work was supported by the Alliance Program of the Helmholtz Association
(HA216/EMMI).

 % do not change 
\end{document}